# Observation of spin-momentum-layer locking in centrosymmetric BiOI

Ke Zhang[1*], Shixuan Zhao[2*], Zhanyang Hao[2*], Shiv Kumar[3], Eike. F. Schwier[3], Yingjie Zhang[2], Hongyi Sun[2], Yuan Wang[2], Yujie Hao[2], Xiaoming Ma[2], Cai Liu[2], Xiaoxiao Wang[3], Koji Miyamoto[3], Taichi Okuda[3], Chang Liu[2], Jiawei Mei[2], Kenya Shimada[3#], Chaoyu Chen[2#] and Qihang Liu[2,4,5#]

[1]*Department of Physical Science, Graduate school of Science, Hiroshima University, Hiroshima, Japan*

[2]*Shenzhen Institute for Quantum Science and Technology and Department of Physics, Southern University of Science and Technology, Shenzhen, China*

[3]*Hiroshima Synchrotron Radiation Center, Hiroshima University, Hiroshima, Japan*

[4]*Guangdong Provincial Key Laboratory for Computational Science and Material Design, Southern University of Science and Technology, Shenzhen 518055, China*

[5]*Shenzhen Key Laboratory of for Advanced Quantum Functional Materials and Devices, Southern University of Science and Technology, Shenzhen 518055, China*

*These authors contributed equally to this work.

#Email: kshimada@hiroshima-u.ac.jp; chency@sustech.edu.cn; liuqh@sustech.edu.cn

## ABSTRACT

Spin polarization effects in nonmagnetic materials are generally believed as an outcome of spin-orbit coupling provided that the global inversion symmetry is lacking, also known as "spin-momentum locking". The recently discovered hidden spin polarization indicates that specific atomic site asymmetry could also induce measurable spin polarization, leading to a paradigm shift to centrosymmetric crystals for potential spintronic applications. Here, combining spin- and angle-resolved photoemission spectroscopy and theoretical calculations, we report distinct spin-layer locking phenomena surrounding different high-symmetry momenta in a centrosymmetric, layered material BiOI. The measured spin is highly polarized along the Brillouin zone boundary, while is almost vanishing around the zone center due to its nonsymmorphic crystal structure. Our work not only demonstrates the existence of hidden spin polarization, but also uncovers the microscopic mechanism of the way spin, momentum and layer locking to each other, shedding lights on the design metrics for future spintronic devices.

*Introduction*

Strategies for generating and controlling highly spin-polarized electronic states in nonmagnetic solids have been explored extensively as a crucial step to realize novel spintronic devices [1-6]. It is generally believed that this requires breaking the space inversion symmetry since a combination of both time-reversal and inversion symmetries inevitably yields the spin-degenerate energy levels. Under such a scenario, a spin splitting induced by spin-orbital coupling (SOC) Hamiltonian [7], is typically classified as Dresselhaus type [8] and Rashba type [9], according to the specific form of inversion symmetry breaking. Recently, new insight pointed out that local symmetry breaking (e.g., polar field) within a part of a unit cell (dubbed as a "sector") can intrinsically lead to a form of "hidden spin polarization" (HSP) in most centrosymmetric crystals [10,11]. While the global inversion symmetry ensures that there is an inversion partner of the given sector manifesting exactly opposite HSP and thus leading to spin-degenerate energy bands in the momentum space, in the real space there are indeed localized spin polarization on each sector [11-13]. The choices of the sector could be a van der Waals layer (e.g., in bulk $WSe_2$ [14]), a sublattice [15], or even an atomic layer (e..g., Se in $PtSe_2$ monolayer [16]), depending on the way of simultaneous symmetry breaking during the observation process. When an individual sector is detected as the majority, the partition of the centrosymmetric unit cell into sectors is naturally selected, say, by the probe, and the corresponding HSP effect can be thus measured by spin- and angle-resolved photoemission spectroscopy (spin-ARPES) [14,16-23] and polarized optical measurement [24-26], etc. Hence, experimental evidences of HSP have been reported based on various layered materials such as bulk and monolayer transition-metal dichalcogenides [14,16-18,20,21], $BaNiS_2$ [19], $LaO_{0.55}F_{0.45}BiS_2$ [22] and Bi2212 cuprate superconductor [23].

Looking for quantum materials with strong HSP effects could considerably expand the material pool for nonmagnetic spintronic device. However, while HSP is simply characterized by the local symmetry breaking in the real space by far, its underlying physics, involving the microscopic mechanism of the way spin, momentum and sector locking to each other, still remains elusive. Recent theoretical work predicts that the magnitude of HSP effect distinguishes a lot around different momenta, such as the center and the boundary of Brillouin zone (BZ) [15]. Here, by using high-resolution spin-ARPES measurements, we investigate the electronic structure and particularly the spin polarization of a single crystal BiOI with nonsymmorphic symmetry. We unambiguously resolve two-fold degeneracy at the Γ point and four-fold degeneracy at the BZ boundary X point, respectively, confirming the nonsymmorphic feature of the bulk band dispersion. More importantly, we observe up to 80% net spin polarization along the BZ boundary (X-M) but almost zero net spin polarization around Γ, indicating a unique momentum dependence of HSP effect. Our tight-binding (TB) model as well as density functional theory (DFT) calculations reveal that in contrast to the Γ point, the nonsymmorphic symmetry helps minimize the spin compensation between different sectors at the BZ boundary, thus successfully retaining the local spin polarization of each sector. Our findings uncover the delicate interplay between spin-momentum locking and symmetry protection in HSP systems, shedding lights on the possibility of all-electrical manipulation for next-generation spintronic devices.

*Electronic structure of bulk BiOI*

BiOI is an ideal semiconductor whose Fermi level is easy to tune by doping, and thus has been extensively studied previously for visible light photocatalysis [27]. BiOI has a tetragonal crystal structure with a centrosymmetric space group P4/nmm containing nonsymmorphic operations of glide mirror $\{M_z|(\frac{1}{2},\frac{1}{2},0)\}$, screw axis $\{C_{2x}|(\frac{1}{2},0,0)\}$ and $\{C_{2y}|(0,\frac{1}{2},0)\}$. The inversion center locates in the middle of two inequivalent O atoms (site point group $D_{2d}$), while the Bi and I atoms occupy the noncentrosymmetric polar sites with the site point group $C_{4v}$. The polyhedrons coordinated by Bi and I atoms are intersected by the O plane. Hence, the quasi-2D unit cell is divided into two sectors $\alpha$ and $\beta$, which are connected by the space inversion operation [see Fig. 1(a)]. The global centrosymmetric structure creates opposite local polar fields along the $c$ axis felt by each BiI layer, which is a prerequisite for the HSP effect.

The Brillouin zone (BZ) and DFT calculated electronic structure of BiOI with SOC are shown in Fig. 1(b) and 1(c), respectively (see Supplementary for computational and experimental methods [28]). The Fermi level is set at the valence band maximum (VBM), which is close to the X point. It is noticeable that at the X and M points, the glide reflection symmetry $\{M_z|(\frac{1}{2},\frac{1}{2},0)\}$ anticommutes with the inversion operator, leading to an extra two-fold degeneracy between two pairs of Kramer's degeneracy, i.e., four-fold degeneracy including spin [29]. Such four-fold degeneracy maintains along the entire X-M line in the absence of SOC [28]. Thus, the band splitting along X-M shown in Fig. 1(c) is totally induced by SOC. In analogy to the conventional Rashba/Dresselhaus effect, such splitting is indeed two sets of spin splitting from sector $\alpha$ and $\beta$ overlapping with each other [15]. In comparison, the splitting along Γ-X is contributed by both orbital repulsion and SOC effect, and is thus larger than that along X-M. The orbital projection analysis shows that in the vicinity of the Γ and X points, the top two valence bands (designated as VB1 and VB2) are mainly composed by $p_x + p_y$ and $s$ orbitals of iodine, while VB3-VB6 are dominated by $p_z$ and $s$ orbitals.

Consistent with its quasi-2D feature of the crystal structure, the electronic structures of BiOI from both DFT calculation and ARPES measurement show generally 2D behavior with relatively flat dispersion along the $c$ axis. The ARPES results measured at a photon energy of 65 eV are shown in Fig. 1(d) [constant energy contours (CECs)], Fig. 1(e-f) (band dispersions) and Fig. 1(g-h) [energy-distribution curves (EDCs)]. From our systematic photon energy dependent measurement [28], this photon energy covers the 6th bulk Γ point. A square-like CEC exists at –1.3 eV, whose corners are locating at X points. As the energy is lowered, the CEC features at X point expand and eventually form contours surrounding M point and merge with those centered at Γ point. This hole-like behavior is clearly presented in the ARPES spectra along M-X-M line in Fig. 1(e). From the CECs and spectra results, we find that the VBM is located around the bulk X point, ~1.4 eV below the experimental Fermi level.

By directly comparing the calculated bulk band structure with the ARPES data shown in Fig. 1(e) and 1(f), one can find good agreement between them, indicating that the surface effect that breaks the global inversion symmetry is rather weak. The predicted four-fold degeneracy at X and M points, and the splitting two-fold degenerate branches (VB1 to VB6) away from X and M are all supported by the ARPES measured dispersion. Furthermore, Fig. 1(g) and 1(h) show the EDCs measured along M-X-M and X-Γ-X directions, more clearly revealing the energy band dispersion details. Especially, at the X point, degenerate peaks, *i.e.*, $X_{1,2}, X_{3,4}$ and $X_{5,6}$ are unambiguously present, while at the Γ point, each of them splits into two peaks, *i.e.*, $\Gamma_1$ to $\Gamma_6$. Consequently, three pairs of Rashba-like hole-type valence bands are formed at the X and M point with the band crossing points locating around -1.4 eV, -2.1 eV and -3 eV for X point, respectively [Fig. 1(e)]. This is in great consistency with our theoretical prediction that only the time-reversal invariant momenta at the BZ boundary (e.g., the X point) possess four-fold degeneracy that is favorable for hidden Rashba effect, while the Γ point does not. We next use spin-ARPES measurements to further demonstrate that the HSP effects surrounding these two high-symmetry points remarkably distinguish with each other.

*Hidden spin polarization*

Figure 2 present the in-plane spin polarization of BiOI measured by spin-ARPES using photon energies of 65 eV for panels (a-b) and 30 eV for panels (c-f), respectively. The spin polarized EDCs from horizontal X-M direction, vertical Γ-X direction, and horizontal Γ-X direction are measured. The comprehensive measurement involving different photon energies and geometries verifies that the observed spin polarization is intrinsic, not affected by the geometrical configurations or final state effect. The representative spin EDCs for the three pairs of two-fold degenerate bands VB1-VB6 are shown in Fig. 2(b), Fig. 2(d) and Fig. 2(f), with the upper (lower) row showing the spin-resolved EDCs and the corresponding $S_y$ ($S_x$) spin component. At the three time-reversal invariant points M, X and Γ (momentum points ①, ③ and ⑩), the spin-resolved EDCs overlap with each other, indicating negligible spin polarization. This is consistent with the spin degeneracy originated from Kramer's pairs.

When the momenta move away from the X point, we observe significant spin polarization as strong as 80% along both $k_x$ and $k_y$ directions (momentum points ②, ④, ⑤, ⑥ and ⑫). Especially, for momenta ⑤ and ⑥, nearly all the six VBs can be well resolved as individual polarization peaks with opposite polarization signs in each pair. This is because the band splitting along Γ-X is more significant, say, if compared with that of X-M [see Fig. 1(c), 1(e) and 1(f)]. In sharp contrast, the spin polarization surrounding the Γ point is very weak (<50% for momentum points ⑦, ⑧, ⑨ and ⑪), especially for the vertical Γ-X direction. Note that one has to make a trade-off between the efficiency and resolution in spin-ARPES measurement. The current resolution configuration may miss the exact high symmetry points in momentum and result into the residual spin polarization signal in ⑩, indicating that the real spin polarization around Γ might be even smaller than the measured value.

Due to the limited photoelectron escape depth ~5 Å [30] and a large lattice constant c = 9.12 Å [31], the photoemission signal mainly comes from the topmost sector (Sector $\alpha$) of the cleaved BiOI single crystal, which is favourable to detect the spin polarization from a local sector. Compared with the previous measurements of HSP materials such as WSe$_2$ [14], PtSe$_2$ [16], LaO$_{0.55}$F$_{0.45}$BiS$_2$ [22] and Bi2212 [23], our work unprecedently measured the distinct polarization features surrounding different high-symmetry points, i.e., BZ center (Γ) and BZ boundary (X), and observed sharp contrast between them. Such observations not only suggest that momentum-dependent spin polarization originates from HSP rather than simply from surface potential gradient, but also imply the key ingredients affecting HSP effect, such as nonsymmorphic symmetry and orbital characters.

*Spin-momentum-layer locking*

In addition to the momentum dependence and high magnitude at the X point, another feature of the HSP in BiOI is the spin texture localized on the measured sector, manifesting a novel way of spin-momentum-layer locking [32-34]. As shown in Fig. 2, for horizontal M-X and horizontal Γ-X lines, $S_y$ component is strong while $S_x$ vanishes. Similarly, for vertical Γ-X line [Fig. 2(c) and 2(d)], $S_x$ component is strong while $S_y$ vanishes. These finding indicates a perpendicular spin orientation to the wavevector. We further confirm the specific spin texture for all the three pairs of valance bands (VB), as illustrate in Fig. 3(a) and 3(b). Note that the spin textures of VB1, VB3 and VB5 are shown in Fig. 3(b), while VB2, VB4 and VB6 have opposite spin patterns, respectively. Surprisingly, while VB1-2 pair shows rather small spin polarization, VB3-4 and VB5-6 pairs exhibit Dresselhaus-type spin textures with large magnitude, rather than Rashba spin polarization induced by the local polar field.

Since the intensity of the incident light decay exponentially with the penetration depth, sector $\alpha$ dominantly contributes the emergent electrons. In addition, according to the photon polarization selection rules [35,36], a *p*-polarized incident light with a specific glancing angle hitting the material surface will leads excitation of electrons in $p_z$ and another in-plane *p*-orbital, depending on the azimuth angle of sample. Therefore, we perform DFT calculations on the $p_z$-orbital projected spin textures localized on the top BiI layer, i.e., sector $\alpha$, as shown in Fig. 3(c) and 3(d), and find excellent agreement with the counterparts measured by spin-ARPES. All of three VB pairs exhibit weak spin polarization ($< 20\%$) around the $\Gamma$ point [28]. As shown in Fig. 3(d), in the vicinity of X, only VB1-2 pair manifests very weak spin polarization, due to the little contribution of $p_z$ orbital of these bands. In comparison, the spin textures of VB3-4 and VB5-6 around X have Dresselhaus type with considerable magnitude. Moreover, the spin patterns of VB3 and VB5 are opposite to each other, which also agrees with the experiment. Thus, we conclude that the DFT results successfully reproduce the features of the experimental observation, strongly supporting that measured spin polarization originates from the intrinsic HSP in BiOI.

To further understand the mechanism of the momentum-dependent HSP effect and the corresponding spin-momentum locking, especially the Dresselhaus spin textures at the X point, we next construct a single-orbital tight-binding model of a nonsymmorphic P4/nmm structure. Two $p_z$ orbitals of the iodine atoms $\alpha$ and $\beta$, connected by the glide mirror operation $\{M_z|(\frac{1}{2},\frac{1}{2},0)\}$, are chosen, as shown in Fig. 3(c). Under the basis of $\{|\alpha \uparrow\rangle, |\alpha \downarrow\rangle, |\beta \uparrow\rangle, |\beta \downarrow\rangle\}$, the model Hamiltonian reads [15]:

$$H(k) = t_1 \cos\frac{k_x}{2} \cos\frac{k_y}{2} \tau_x \otimes \sigma_0 + t_2(\cos k_x + \cos k_y)\tau_0 \otimes \sigma_0 + \lambda\tau_z \otimes (\sigma_y \sin k_x - \sigma_x \sin k_y), (1)$$

where $\tau$ and $\sigma$ are Pauli matrices under the basis of $\{|\alpha\rangle, |\beta\rangle\}$ and $\{|\uparrow\rangle, |\downarrow\rangle\}$, respectively; $t_1$ and $t_2$ present inter-sector and intra-sector electron hopping, contributing to diagonal and off-diagonal terms of the Hamiltonian, respectively. The third term of Eq. (1) describes the SOC effect induced by the local symmetry breaking for each sector, parametrized by $\lambda$. It is noticeable that at the boundary of Brillouin zone, e.g., the X-M line, Eq. (1) naturally becomes block diagonal for $\{|\alpha\uparrow\rangle, |\alpha\downarrow\rangle\}$ and $\{|\beta\uparrow\rangle, |\beta\downarrow\rangle\}$ with opposite local spin polarization for each sector. When the probe sees sector $\alpha$ predominately, i.e., breaking the symmetry between $\alpha$ and $\beta$, Eq. (1) is naturally decomposed into two matrices for each sector. Thus, the HSP of sector $\alpha$ is sizable enough to be measured. In contrast, the HSP effect is remarkably suppressed around the Γ point because of the inter-sector coupling $t_1$ term, as shown in Fig. 4(e), leading to negligible spin signal from spin-ARPES.

Our tight-binding model also helps understand the specific spin textures around different high-symmetry momenta. The low-energy effective $k \cdot p$ Hamiltonians derived from Eq. (1) take the form of $(k_x\sigma_y - k_y\sigma_x)\tau_z$ at M and $(k_x\sigma_y + k_y\sigma_x)\tau_z$ at X, indicating Rashba and Dresselhaus type HSP, respectively, the latter of which perfectly explains the measured spin polarization around the X point. Such results indicate that although a (local) polar field existing in a crystal in general supports a (hidden) Rashba-type spin polarization, a (hidden) Dresselhaus-type spin polarization would also be accompanied [11], depending on the specific symmetry of a given momenta.

In summary, combining spin-ARPES measurements and theoretical calculations, we unprecedently report distinct spin-momentum-layer locking phenomena at different position of the BZ in a centrosymmetric material BiOI. The measured spin polarization localized on a specific BiI layer is highly polarized along the BZ boundary but almost vanishing around the zone center, due to its nonsymmorphic crystal structure. In addition, the pattern of the layer-resolved spin texture, either Rashba or Dresselhaus type, also reflects the symmetry of both real space and *k*-space. Our finding not only experimentally demonstrates the existence of HSP effect, but also sheds light on the design metrics for significant spin polarization in centrosymmetric materials by uncovering the intimate interplay between spin, orbital and layer degrees of freedom.

**Acknowledgments**

This work is supported by the National Natural Science Foundation of China (NSFC) (Grant Nos. 11874195 and 12074163), the Shenzhen High-level Special Fund (Grant Nos. G02206304, G02206404), the Guangdong Innovative and Entrepreneurial Research Team Program (Grant No. 2019ZT08C044), the Shenzhen Science and Technology Program (Grant No.KQTD20190929173815000), the Innovative Team of High School in Guangdong Province (Grant No. 2020KCXTD001), Guangdong Provincial Key Laboratory for Computational Science and Material Design under Grant No. 2019B030301001 and Center for Computational Science and Engineering of Southern University of Science and Technology. ARPES measurements were performed with the approval of the Proposal Assessing Committee of the Hiroshima Synchrotron Radiation Center (Proposal Nos. 19AG004, 19BG014). We would also like to thank the N-BARD, Hiroshima University for supplying liquid He.

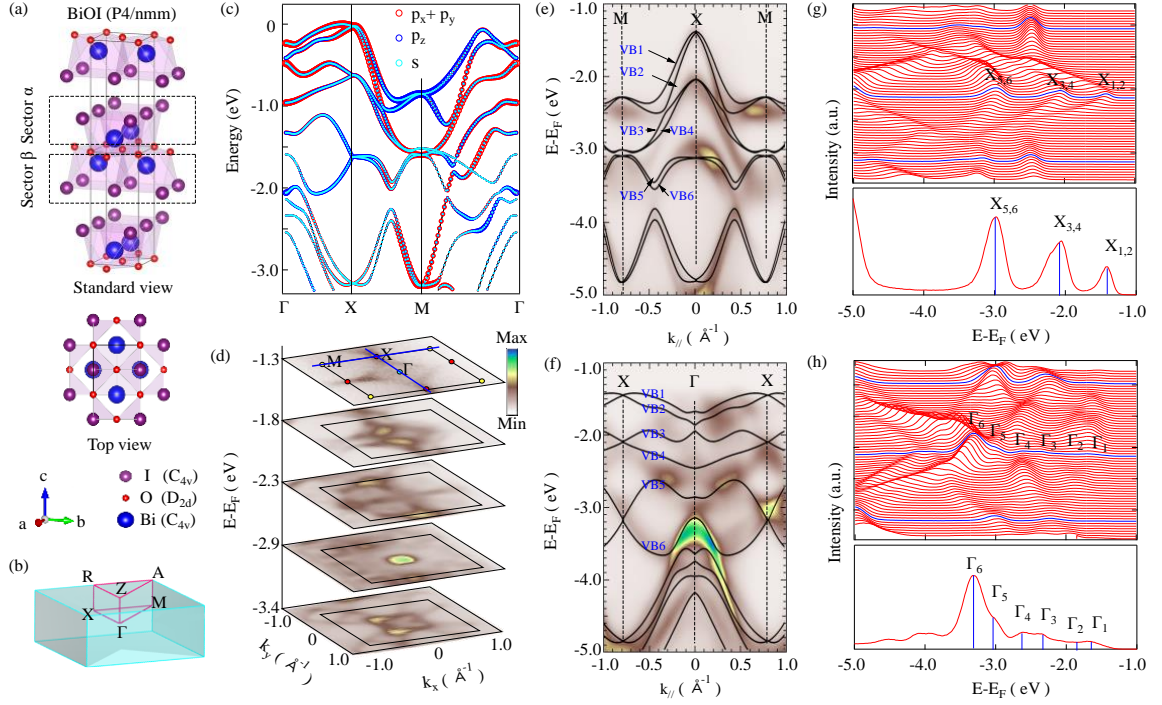

**Fig. 1: Crystal and electronic structure of BiOI.** (a) Standard and top views of the single crystal structure of BiOI. The unit cell consists of two BiI layers as inversion partners, labelled as sector $\alpha$ and $\beta$. (b) Corresponding bulk Brillouin zone. (c) DFT calculated bulk band dispersion with orbital projection. (d) ARPES measured CECs of the valence bands at different energies. (e), (f) ARPES measured spectra along the X-M and $\Gamma$-X high symmetry lines, overlaid by DFT calculated dispersions (black solid lines). (g), (h) EDCs corresponding to the spectra shown in (e) and (f), respectively. The lower parts are the EDCs at X and $\Gamma$ points, respectively, from which one can resolve the spectral peaks corresponding to the top six valence bands.

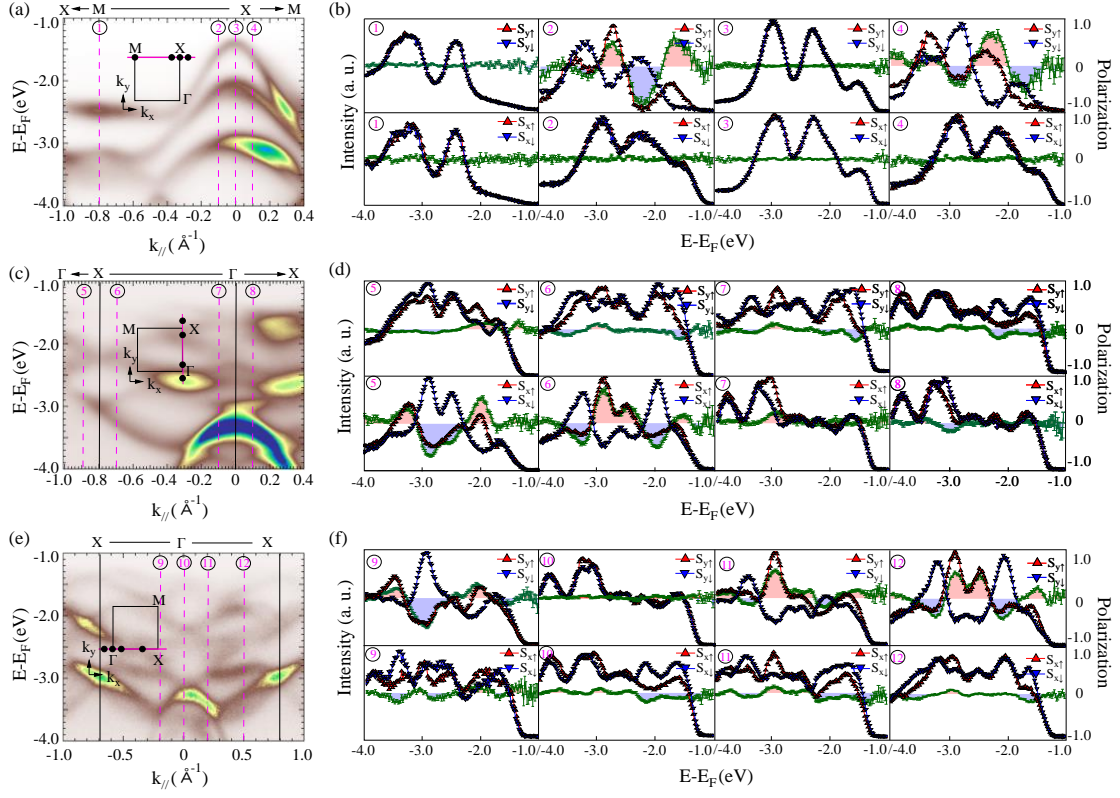

**Fig. 2: Hidden spin polarization observed around the X and Γ points.** (a), (c), (e) Band dispersion along the top M-X, vertical Γ-X and horizontal Γ-X directions, respectively. For each panel, the inset shows the BZ with black dots indicating the momentum positions where the spin-resolved EDCs are taken. (b), (d), (f) Spin-resolved EDCs measured using spin-ARPES and corresponding spin polarizations. The number for each panel corresponds to the momentum point denoted by the pink dashed lines in (a), (c) and (e). The data is taken at 30 K shown with the photo energies $hv = 65$ eV for (a) and (b), and $hv = 30$ eV for (c-f).

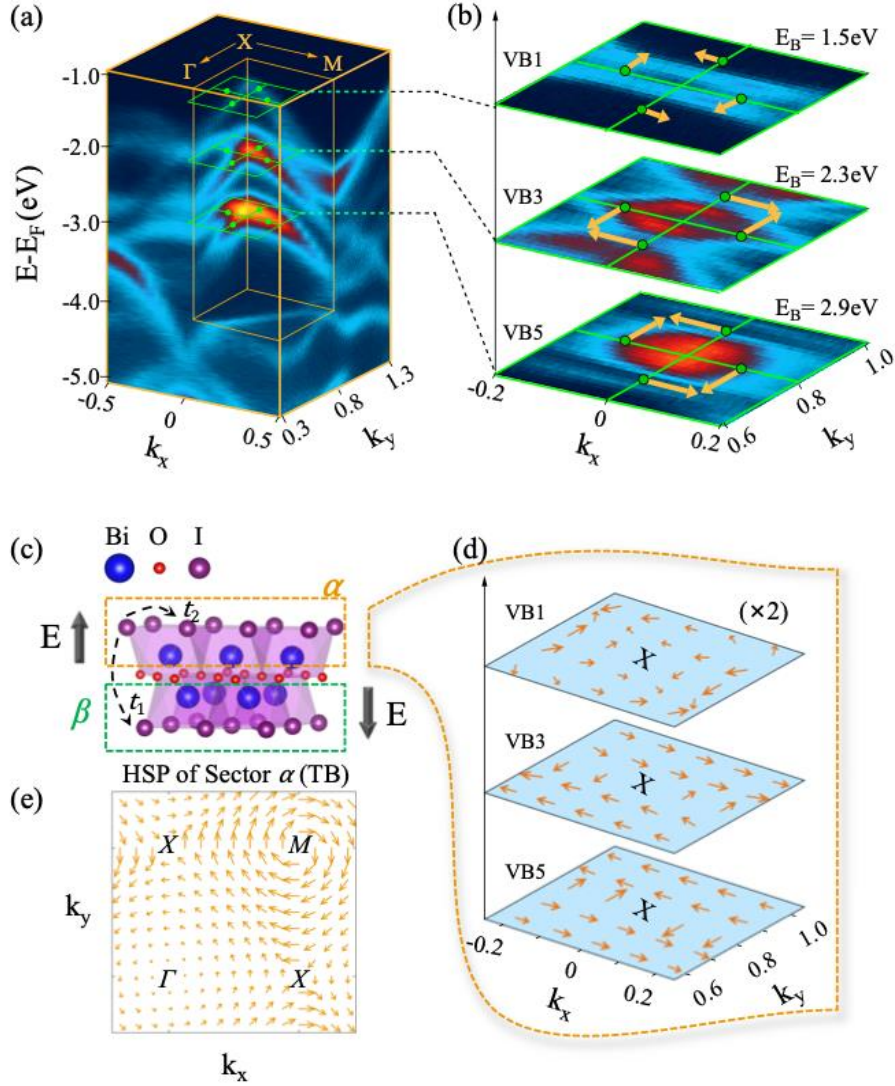

**Fig. 3: Spin-momentum-layer locking in BiOI.** (a) Overview of ARPES measured band dispersion of BiOI. (b) Schematic sketch of the measured spin textures of VB1, VB3 and VB5 by spin-ARPES, with the momentum cross section denoted by the green squares in panel (a). (c) Layered structure of BiOI with two BiI sectors feeling opposite local dipole fields (black arrows). (d) DFT calculated $p_z$-projected HSP of VB1, VB3 and VB5 around X for sector $\alpha$. The spin magnitude of VB1 is multiplied by two. (e) Spin texture for sector $\alpha$ calculated by our tight binding (TB) model, showing Dresselhaus and Rashba type HSP effect for X and M, respectively.